\documentclass[atmp]{ipart_v1} 
\firstpage{01}
\Year{To appear in volume 20 (2016)}   

\newcommand{\beq}{\begin{equation}}
\newcommand{\eeq}{\end{equation}}
\newcommand{\beqa}{\begin{eqnarray}}
\newcommand{\eeqa}{\end{eqnarray}}
\newcommand{\nn}{\nonumber}

\newcommand{\half}{\frac{1}{2}}

\newcommand{\pbr}[2]{ \{ \hspace*{-2.2pt} [ #1 , #2\hspace*{1.5 pt} ]
\hspace*{-2.7pt} \} }
\newcommand{\we}{\wedge}

\newcommand{\der}{\partial}

\newcommand{\om}{\omega}

\newcommand{\inn}{\hspace*{2pt}\raisebox{-1pt}{\rule{6pt}{.3pt}\hspace*
{0pt}\rule{.3pt}{8pt}\hspace*{3pt}}}

\newcommand{\ka}{\varkappa}

\newcommand{\al}{\alpha}

\newcommand{\deltab}{\bar \delta}

\newcommand{\what}[1]{\widehat{#1}}

\newcommand{\bx}{{\bf x}}
\newcommand{\bk}{{\bf k}}

\newcommand{\BPsi}{{\bf \Psi}}
\newcommand{\BPhi}{{\bf \Phi}}
\newcommand{\BXi}{{\bf \Xi}}
\newcommand{\BH}{{\bf H}}
\newcommand{\BS}{{\bf S}}

\newcommand{\rd}{\mathrm{d}} 

\numberwithin{equation}{section}

\begin{document}

\author[Igor V. Kanatchikov]{Igor V. Kanatchikov\vskip1em   
}

\title[Precanonical structure of $\cdots$]{On the precanonical structure of the Schr\"odinger wave functional}

\begin{abstract}
We show that the Schr\"odinger wave functional may be obtained as the product integral of precanonical wave functions on the space of
field and space-time variables.
 The functional derivative Schr\"o\-din\-ger equation
 underlying the canonical field quantization is
 derived from the partial derivative covariant analogue
 of the Schr\"odinger equation, which  appears 
 in the precanonical field quantization \linebreak 
 based on the De Donder-Weyl generalization of the Hamiltonian formalism for field theory.
The representations of precanonical quantum operators typically contain 
an ultraviolet parameter $\varkappa$ of the dimension 
of the inverse spatial volume. 
 The transition from
the precanonical description of quantum fields in terms of Clifford-valued wave functions
and partial derivative operators
to the standard functional Schr\"odinger representation obtained from canonical quantization is accomplished if 
$\frac{1}{\varkappa}\rightarrow 0$
and  $\frac{1}{\varkappa}\gamma_0$ is mapped to the infinitesimal spatial volume element $\rd\mathbf{x}$. Thus the standard QFT 
obtained via canonical quantization 
 corresponds to the quantum theory of fields 
 derived 
  via 
  precanonical quantization in the limiting case of
 an infinitesimal value of the parameter $\frac{1}{\varkappa}$.
\end{abstract}

\maketitle

\section{Introduction }

Field theories are commonly considered as systems with an infinite number of degrees of freedom. This notion originates in the canonical
Hamiltonian treatment of field theory and it is transferred to
quantum field theory by the procedure of canonical quantization.
The resulting version of quantum field theory
has evolved  into a 
very successful framework in contemporary theoretical physics whose
applications range from condensed matter physics to quantum
cosmology. In this framework, even the divergences, viewed as pathologies in the earlier days of QFT, have turned into triumphs for the concepts of normalization and effective field theory.
 
However, there remain conceptual tensions between quantum theory
and relativity,
which we face in the context of discussions of foundational issues, and particularly in 
quantum field theory in curved space-times, quantum gravity, and unification of all interactions,  where our current understanding of QFT is  pushed to the limits of its applicability due to the distinguished role of time in the formalism of quantum theory on the one side and the generally covariant,  geometric and nonlinear nature of general relativity on the other.

 In approaching those issues, we draw attention to the fact that the progress of QFT has essentially overlooked developments in the calculus of variations of multiple integral problems,  where the extension of the Hamiltonian formulation from mechanics to field theory is known to be far from unique \cite{kastrup}. 
Moreover, as opposed to the canonical Hamiltonian formalism,
its generalizations developed in the setting 
of the calculus of variations have no need of a distinguished "time variable" in 
the set of space-time variables (i.e. the independent variables of the variational problem  
which defines a field theory). 
 Nor do  they necessarily imply the picture of fields as infinite-dimensional systems evolving in time (which would fail on non-globally hyperbolic space-times).

 Thus, the question arises
whether  a formulation of
quantum field theory could be built based on these alternative
space-time symmetric Hamiltonizations of field theory,
 and if the inherent features of the latter, such as
manifest respect for the space-time symmetries and 
 the 
 finite dimensionality
of the corresponding analogue of the configuration 
space (i.e. the bundle of field variables
over space-time, whose sections are field configurations appearing in the standard formulations) can help in clarifying fundamental
issues of QFT at the 
frontiers of current research, e.g. in the context of quantum gravity.

Furthermore, the existence of the Hamilton-Jacobi formulations of field theories
associated with each of these alternative Hamiltonizations 
 \cite{kastrup,rund,nambu,horava,my-epl} 
naturally leads to the question whether 
alternative formulations 
of quantum field theories exist which would reproduce the corresponding 
generalized 
Hamilton-Jacobi equations in  the classical limit,
and what would be their physical significance.

To be more specific, let us recall that in a 
field theory given by
the first order Lagrangian density $L=L(y^a,  y^a_\mu, x^\mu)$,
where $y^a$ denote
 the 
field variables of any nature, $y^a_\mu$ are
(the first jet space coordinates of) their first derivatives,
 and  $x^\mu$ are
 the 
space-time variables,
the simplest of the above mentioned alternative Hamiltonizations
is the so-called De Donder-Weyl (DW) theory (see e.g. \cite{kastrup,rund}).
 It  is based on the following Hamiltonian-like covariant reformulation of the
Euler-Lagrange field equations:
\beq \label{dw}
\der_\mu p^\mu_a = - \frac{\der H}{\der y^a},
\quad \der_\mu y^a = \frac{\der H}{ \der p^\mu_a}~,
\eeq
which uses the following 
covariant Legendre transformation to define new variables:
$p^\mu_a :=\der L/\der y^a_\mu $ ({\em polymomenta}) and
$H (y^a,p^\mu_a,x) := p^\mu_a\der_\mu y^a -L $  ({\em DW Hamiltonian function}).

The DW Hamiltonian equations (\ref{dw}) can be compared with the standard Hamilton
equations in the canonical formalism:
\beq \label{canon}
\der_t p_a^0(\bx)=-\frac{\delta \BH}{\delta y^a(\bx)}, \quad
\der_t y^a(\bx) = \frac{\delta \BH}{\delta p^0_a(\bx)}~,
\eeq
where the canonical Hamiltonian functional is introduced:
\beq
\BH ([y(\bx),p^0(\bx)]) := \int\! \rd\bx\, \Big(\der_t y^a(\bx) p_a^0(\bx) -
L \Big)~,
\eeq
and a decomposition into
the
space and time is performed, so that
$x^\mu := (\bx, t)$.
 Here and in what follows the capital bold letters denote functionals.

When both formulations are regular
the equivalence between (\ref{dw}) and (\ref{canon})
can be established  by noticing that
\beq
\BH = \int\! \rd\bx\, \Big ( H-\der_i y^a(\bx) p_a^i(\bx)\Big)~.
\eeq
Then it is easy to check that
 the 
 canonical Hamilton equations can be derived from the
(precanonical) DW Hamiltonian equations
(\ref{dw}).

Whereas the field quantization based on the canonical Hamiltonization
is well elaborated and underlies QFT as we know it,
an approach to quantization of
fields based on the De Donder-Weyl (DW) generalization of
Hamiltonian mechanics to field theory
was put forward
 only recently
\cite{ka-qft,ka-qft2,ka-qft3}  (c.f. also discussions of similar ideas by other authors in
\cite{norris,navarro,sardan-q,hamburg,nik-q}).
In the context of quantization of gravity \cite{ik2,ik4,ik5}
the approach was later given the name of
{\em precanonical quantization.}

While  the connection between 
 the canonical and DW Hamiltonizations is sufficiently
clear on the classical level (see e.g. \cite{gimmsy}),
the relation between the respective quantizations has been
rather problematic for a long time (see \cite{ka-qft,ik-pla} for earlier discussions).
In the recent paper \cite{myschrod11} we found a formula connecting
the Schr\"odinger wave functional with the precanonical wave functions in the case of
scalar field theory. However, the derivation was based on an 
{\em ad hoc\/}
Ansatz, so that it remained unclear
 how general the result is.
 In this paper we establish a connection
between QFT based on canonical quantization in the functional 
Schr\"odinger representation
and a 
 formulation based on precanonical quantization 
 without any {\em a  priori\/} assumptions
regarding the form of this relation,  except 
 a very general possible relation between the Schr\"odinger 
wave functional and precanonical wave function 
(c.f. Eq. (\ref{bpsi}) below).  

In Sect. 2 we  present a comparative outline of the elements of canonical and precanonical quantization, which are essential for our purposes.
In Sect. 3 we  derive the functional derivative Schr\"odinger equation
for quantum scalar field theory from the corresponding Dirac-like partial
derivative precanonical analogue of the Schr\"odinger equation.
This consideration leads to a relation between the Schr\"odinger wave functional
  known from the 
  canonical quantization and the Clifford-valued wave function
  appearing in 
  precanonical quantization. As an application of our result,
   we construct the
vacuum state functional of the free scalar field theory from the
precanonical ground state wave functions. A 
 concluding discussion is found in  Sect. 4.

\section{Canonical and precanonical quantization}

Let us present a brief comparative overview of
the elements of canonical and precanonical quantization, which are
relevant for the  following discussion.

Canonical quantization  (in the Schr\"odinger picture \cite{hatfield})
is known to lead
to the description of quantum fields in terms of the Schr\"odinger wave
functional $\BPsi([y(\bx)],t)$ on the infinite-dimensional configuration space
of field configurations $y(\bx)$ at time $t$.
Precanonical quantization \cite{ka-qft,ka-qft2,ka-qft3}   leads to the description in terms of
Clifford algebra-valued wave
functions $\Psi(y,x)$ on the finite dimensional ``covariant configuration space"
(in the terminology of \cite{gimmsy})
of field variables $y$ and space-time variables $x$.

While  the Schr\"odinger wave
functional $\BPsi$ fulfils  the Schr\"odinger equation
\beq \label{se}
i\hbar\der_t\BPsi = \what{\BH} \BPsi~,
\eeq
where $\what{\BH}$ stands for the functional derivative operator of the canonical Hamiltonian,
the precanonical wave function
$\Psi(y,x)$ satisfies the following covariant generalization
of the Schr\"odinger equation \cite{ka-qft,ka-qft2,ka-qft3}
\beq \label{nse}
i\hbar\ka\gamma^\mu\der_\mu \Psi = \what{H}\Psi~,
\eeq
where $\what{H}$ is the partial derivative operator of the De Donder-Weyl Hamiltonian function,
$\gamma^\mu$ are Dirac matrices of  $n$-dimensional space-time,
and $\ka$ is a ``very large'' constant of dimension $\ell^{-(n-1)}$.
 The latter routinely appears on dimensional grounds in the
 expressions of 
precanonical quantum operators, which follow from quantization of
the Poisson-Gerstenhaber brackets of differential forms representing
the dynamical variables in field theory.  These brackets  were found in our
 earlier work on the mathematical structure of the DW Hamiltonian formulation  \cite{ikanat,bial96,go96} 
(for further generalizations see e.g.
\cite{paufler,baez,my-dirac,mirco,tamar,vey}).
 Their geometric prequantization, 
which can be viewed as an intermediate step toward precanonical quantization,
was considered in \cite{gq}. It is on the level of geometric prequantization 
we can justify the appearance of Clifford algebra-valued wave functions and 
operators in precanonical quantization.

More specifically, let us consider the theory of interacting scalar fields,
 which is given by
\beq
L=\half\der_\mu y^a \der^\mu y^a - V(y)~,
\eeq
where the potential term $V(y)$ also includes the mass terms like $\frac12 m^2y^2$
(henceforth we set $\hbar=1$ and use the metric signature $+-...-$).

 In this case
the operator of
 the 
 canonical conjugate momentum of $y^a(\bx)$:
$${p}{}_a^0(\bx):=\frac{\der L}{\der \der_t y^a(\bx)}~,$$
in the Schr\"odinger $y(\bx)$-representation
 is given by
\beq
\hat{p}{}_a^0(\bx) = -i 
\frac{\delta}{\delta y^a(\bx)}~.
\eeq
This representation follows from quantization of the equal-time Poisson bracket
\beq
\{ p_a^0(\bx),y^b(\bx') \}=\delta^b_a \delta(\bx-\bx').
\eeq

In precanonical quantization,
the representation of the operators of polymomenta:
\beq
\hat{p}_a^\mu = 
- i \ka\gamma^\mu \frac{\der}{\der y^a}~,
\eeq
follows from quantization of the Heisenberg subalgebra
of the above mentioned
Poisson-Gerstenhaber algebra of forms 
(c.f. \cite{ka-qft,ka-qft2,ka-qft3,my-ehrenfest}):
\beqa
&&\pbr{p_a^\mu\omega_\mu}{y^b}
=
\delta^b_a , \quad \nn \\
&&\pbr{p_a^\mu\omega_\mu}{y^b\omega_\nu}
=
\delta^b_a\omega_\nu, \quad \\
&&\pbr{p_a^\mu}{y^b\omega_\nu}
=
\delta^b_a\delta^\mu_\nu ,    \nn
\eeqa
where $\om_\mu := \der_\mu\inn\omega$ are the contractions 
of the volume $n$-form  on the space-time 
\mbox{$\omega:=\rd x^0\we \rd x^1\we...\we \rd x^{n-1}$}
with the basis vectors $\der_\mu$ of its tangent space.
 This quantization
also implies the existence of a  map $q$  from  the co-exterior
forms on the classical level\footnote{We have explained in our
earlier papers \cite{bial96}
that the natural multiplication of forms here is given by the
 {\em co-exterior } product: $\al\bullet\beta:= *^{-1}(*\al\wedge *\beta)$,
where $*$ is the Hodge duality operator.
}
to the Clifford algebra elements (Dirac matrices) on the quantum level:
\beq \label{qmap}
\om_\mu \stackrel{q}{\longmapsto}
\frac{1}{\ka}\,\gamma_\mu ,
\eeq
which is similar to the
``Chevalley map" \cite{chevalley},
or ``quantization map" \cite{mein},  known in  the theory of Clifford algebras.
 The constant $\ka$ appears here on dimensional grounds.
  From the association of $\omega_0$, which represents the
infinitesimal spatial volume element $\rd\bx$,
 with
 $\gamma_0$, which is dimensionless, it is evident that
   $\frac{1}{\ka}$ corresponds to a ``very small"
volume and has
the
meaning of a physically infinitesimal or elementary volume.

Furthermore, while the  canonical Hamiltonian operator of the quantum scalar field theory:
\beq \label{hse}
\what{\BH}= \int\! \rd\bx \left \{
-\half\frac{\delta^2}{\delta y(\bx)^2} + \half (\nabla y(\bx))^2 + V(y(\bx))
\right \} ,
\eeq
is formulated in terms of functional derivative operators,
the DW Hamiltonian operator in this case (see \cite{ka-qft,ka-qft2,ka-qft3}):
\beq \label{hscalar}
\what{H}= -\half \ka^2 \der_{yy} + V(y)
\eeq 
 is expressed in terms of 
 the partial derivative operators with respect to the field variables.

The question naturally arises,  how those two descriptions, which seem
to be so different both physically and mathematically,
 can be related: how the Schr\"odinger wave functional is related 
 to the Clifford algebra-valued
 precanonical wave function and how the functional derivative canonical 
 Schr\"odinger equation (\ref{se}) is related to the
 precanonical Schr\"odinger equation (\ref{nse}).

\section{Schr\"odinger wave functional from  precanonical wave function}

 To make the  above mentioned relation less mysterious,
  let us first recall our earlier observation \cite{ik-pla}
  that the functional derivative Hamilton-Jacobi equation of the canonical Hamiltonian
formalism:
\beq \label{bh}
\der_t \BS +
\BH\left (y^a(\bx), p^0_a(\bx)=\frac{\delta \BS}{\delta y(\bx)},t \right )=0 ,
\eeq
can be derived from
the partial differential Hamilton-Jacobi equation
of precanonical De Donder-Weyl theory \cite{kastrup,rund}:
\beq
\der_\mu S^\mu + H\left (y^a,p^\mu_a=\frac{\der S^\mu}{\der y^a},x^\mu\right )=0 ,
\eeq
if the canonical HJ functional
$\BS([y(\bx),t])$ is constructed  in terms of the DW-HJ functions $S^\mu(y,x)$
as follows: %
\beq \label{dwhjs}
\BS = 
  \int_\Sigma\!  S^\mu \omega_\mu .
\eeq
Here $\Sigma$: ($y\!=\!y(\bx), t\!=\!const$) is the subspace in the covariant
configuration space,  which represents the field configuration $y(\bx)$ at the moment~of~time~$t$.

This result of \cite{ik-pla} demonstrates that
the transition from
 an object of DW (precanonical) theory, such us
$S^\mu(y,x)$, to
 an object of canonical theory, such as $\BS([y(\bx),t])$, involves
a restriction of the former to
the  
 subspace $\Sigma$ 
and subsequent integration over it.
In this way the functionals of field configurations are constructed from
the functions on the covariant finite-dimensional configuration space,
and the functional derivative equations of the canonical formalism
are derived from their
precanonical partial derivative counterparts. 

A similar relationship exists also between the forms representing 
 dynamical variables and their Poisson-Gerstenhaber brackets 
 in the DW Hamiltonian  formulation and 
 the ``observables" represented by functionals 
 and their  Poisson brackets in the canonical formalism \cite{ikanat,helein}. 

Having obtained this result on the classical level, we can
also
expect a similar relation between the wave functional and precanonical wave function on the quantum level,
because both the Schr\"odinger wave functional
$\BPsi$ and the precanonical wave function $\Psi$ are  related to the exponentials
of, respectively,  the  HJ functional and DW-HJ functions, viz.,
\beq
\BPsi \sim e^{i\BS} \quad \mathrm{and} \quad \Psi \sim e^{\frac{i}{\ka}S^\mu\gamma_\mu}
\eeq
(see \cite{ka-qft3}, where the second expression is used to argue that in the classical limit
 the DW-HJ equation
follows from the precanonical Schr\"odinger equation).
 Using the fact that $\BS$
is the spatial integral of DW-HJ functions,
 we can anticipate that
 $\BPsi$ is related to the {\em product integral } \cite{productintegral}
of precanonical
wave functions restricted to the subspace $\Sigma$:
\beqa \label{pri}
\BPsi([y(\bx)]) \sim e^{i\BS} =
e^{i\!\int_\Sigma\!  S^\mu \omega_\mu}
&=& \prod_{\bx\in \Sigma} e^{i S^\mu(y=y(\bx),\bx,t) \omega_\mu }
\nn \\
&\sim& \prod_{\bx\in\Sigma}
\Psi(y=y(\bx),\bx,t)|_{ \frac{1}{\varkappa}\gamma_\mu\rightarrow \omega_\mu}~,
\eeqa
where the last step implies the inverse of the ``quantization map" in Eq. (\ref{qmap}).
The consideration below  will make this idea more precise.

\medskip

Now, let us assume that the Schr\"odinger wave functional $\BPsi$ can be expressed
in terms of the precanonical wave function $\Psi(y,x)$ restricted to the
  subspace $\Sigma$:
$\Psi(y,x)|_\Sigma:=\Psi_\Sigma (y(\bx), \bx, t)$, i.e.
\beq \label{bpsi}
\BPsi ([y(\bx)], t) = \BPsi ([\Psi_\Sigma (y(\bx), \bx, t)],[y^a(\bx)])~.
\eeq
Then the time evolution of the Schr\"odinger wave functional is
determined by the time evolution of precanonical wave function.
By applying the chain rule to the composite functional (\ref{bpsi})
we obtain:
\beq \label{dtbpsi0}
i\der_t \BPsi =
\int\! \rd\bx~  
{\sf Tr}\left \{
 \frac{\delta \BPsi }{\delta\Psi^T_\Sigma(y^a(\bx),\bx, t)}
i\der_t \Psi_\Sigma (y^a(\bx),\bx,t)
\right \}~.
\eeq
Note that the additional dependence of $\BPsi$ from $y^a(\bx)$, which is not incorporated in
$\Psi|_\Sigma$, is supposed to take into account the fact that
the amplitudes  $\Psi|_\Sigma$ in space-like
separated points in general are  not  independent - 
one of the manifestations of quantum nonlocality. 


The equation of time evolution of  $\Psi_\Sigma (\bx)$ is
given by the space-time decomposed precanonical Schr\"odinger equation,
Eq. (\ref{nse}),  restricted to $\Sigma$, viz.,
\beq  \label{nsesigma}
i \der_t \Psi_\Sigma (\bx) = -i\alpha^i \frac{\rd}{\rd x^i} \Psi_\Sigma  (\bx)
+ i\alpha^i\der_i y(\bx){\der_y} \Psi_\Sigma  (\bx)
+  \frac{1}{\varkappa} \beta (\what{H} \Psi)_\Sigma  (\bx) ,
\eeq
where
\beq\label{tder}
\frac{\rd}{\rd x^i}:= \der_i +\der_i y (\bx)  \der_y + \der_{ij}y (\bx) \der_{y_j} + ...
\eeq
is the total derivative
 along the subspace $\Sigma$,
$\beta:=\gamma^0$, $\beta^2=1$, and $\alpha^i:=\beta\gamma^i$.
 Here we have also introduced
 the shorthand notation $\Psi_\Sigma(\bx) :=\Psi_\Sigma (y(\bx), \bx, t)$ to be used henceforth.

Hence, the time evolution of the wave functional of the quantum scalar field
is given by:
\beqa \label{dtbpsi1}
i\der_t \BPsi 
&\!\!=\!\!&
\int \!\rd\bx\ 
{\sf Tr}\left \{
 \frac{\delta \BPsi }{\delta\Psi^T_\Sigma(\bx, t)}
\left [ -i\alpha^i \frac{\rd}{\rd x^i} \Psi_\Sigma  (\bx)
+ i\alpha^i\der_i y(\bx){\der_y} \Psi_\Sigma  (\bx)
  \right. \right . \nn \\
&&- \left.\left. \frac{1}{2}\varkappa\beta \der_{yy}\Psi_\Sigma  (\bx)+
  \frac{1}{\varkappa}\beta V(y(\bx)) \Psi_\Sigma  (\bx)
\right ] \right \} .
\eeqa
Eq. (\ref{dtbpsi1}) is obtained by inserting the explicit expression of the DW Hamiltonian
operator of the nonlinear scalar field given by Eq. (\ref{hscalar}) into Eq. (\ref{nsesigma}).

 From Eq. (\ref{bpsi}) we obtain
the expressions for the first and the second
total functional derivatives of  $\BPsi$ with respect to $y(\bx)$, viz.,
\beqa \label{delta-bpsi}
 \frac{\delta \BPsi }{\delta y(\bx)}
&=&
{\sf Tr}\left \{
 \frac{\delta \BPsi }{\delta\Psi^T_\Sigma(\bx, t)}
\der_y \Psi_\Sigma (\bx)
\right \}
+  \frac{\deltab \BPsi }{\deltab y(\bx)^{{}^{}}}~, \\
 && \nn \\
  && \nn \\
\frac{\delta^2 \BPsi }{\delta y(\bx)^2}
&=&
{\sf Tr}\left \{
 \frac{\delta \BPsi }{\delta\Psi^T_\Sigma(\bx, t)}
~\delta(\mbox{\bf 0})\der_{yy} \Psi_\Sigma (\bx)
\right \}
 \nn \\
&+&{\sf Tr} \, {\sf Tr} \left \{
 \frac{\delta^2 \BPsi}{\delta \Psi^T_\Sigma(\bx)\otimes\delta\Psi^T_\Sigma(\bx)}
~\der_y \Psi_\Sigma (\bx)
\otimes  \der_y \Psi_\Sigma  (\bx)
\right \} \label{delta-bpsi2} \\
&+& 2~{\sf Tr} \left \{
\frac{\delta \deltab \BPsi}{\delta \Psi^T_\Sigma (\bx)
~\deltab y(\bx)}   ~\der_y \Psi_\Sigma (\bx) \right \}
+  \frac{\deltab^2 \BPsi }{\deltab y(\bx)^2} ~. \nn 
\eeqa
Here and in what follows $\deltab$ denotes
 the 
partial functional derivative
with respect to $y(\bx)$,
and $\delta(\mbox{\bf 0})$
 is the result of functional differentiation
of a function with respect to itself at the same spatial point.

\medskip

{Our {\bf first} observation} is that the potential energy term in the
 canonical functional derivative
 Schr\"odinger equation for the scalar field, see
 Eqs. (\ref{se}) and (\ref{hse}),
should be obtained from
 the potential energy term
 in the precanonical equation, Eq. (\ref{dtbpsi1}),
 i.e.,
\beq
 \int \!\rd\bx\ 
 {\sf Tr}
\left \{
\frac{\delta \BPsi }{\delta \Psi^T_\Sigma (\bx)}
~\frac{1}{\ka}
\beta
V(y(\bx))\Psi_\Sigma (\bx)) \right \}
\mapsto \int \!\rd\bx\ V(y(\bx))~\BPsi~.
\eeq
It can be accomplished if
\beq \label{v-term}
{\sf Tr} \left \{    \frac{\delta \BPsi}{\delta\Psi^T_\Sigma (\bx)}
 ~ \frac{1}{\ka} \beta
 \Psi_\Sigma (\bx) \right \}
  \mapsto \BPsi
\eeq
in  any  point $\bx$.
The precise meaning of the operation $\mapsto $ will be established below.

 \medskip

 The {\bf second} observation is obtained by
 functionally differentiating both sides of Eq. (\ref{v-term}) with respect to
$\Psi^T_\Sigma (\bx)$:
\beq \label{kade}
{\sf Tr} \left \{
\frac{\delta^2 \BPsi}{\delta\Psi^T_\Sigma (\bx)
\otimes \delta\Psi^T_\Sigma (\bx)}
 \frac{1}{\ka} \beta \Psi_\Sigma (\bx)  \right \}
+ \frac{\delta \BPsi}{\delta \Psi^T_\Sigma(\bx)}
 \frac{1}{\ka}\beta \delta(\mbox{\bf 0})
 \mapsto
 \frac{\delta \BPsi}{\delta \Psi^T_\Sigma(\bx)}~,
\eeq
where
$\delta(\mbox{\bf 0}) ={\delta \Psi_\Sigma(\bx)}/{\delta \Psi^T_\Sigma(\bx)}.$
 We conclude, that
 the second term in (\ref{delta-bpsi2}),
 which has no counterparts in the familiar
functional  Schr\"odinger equation, vanishes,  provided
\beq \label{aaa}
\frac{1}{\ka} \beta \delta(\mbox{\bf 0}) \mapsto 1 .
\eeq

\medskip

 Similarly, our {\bf third} observation is that
the term $\ka\beta\der_{yy}\Psi_\Sigma$  in  (\ref{dtbpsi1})
reproduces the  first term in
  Eq. (\ref{delta-bpsi2}) and therefore,
 in the functional Schr\"odinger equation with $\hat{\BH}$ given by Eq.(\ref{hse}),
  if
\beq \label{aaaa}
\beta \ka \mapsto \delta(\mbox{\bf 0}) .
\eeq
We see that  this condition is consistent with Eq. (\ref{aaa})
in the sense that (\ref{aaa}) is fulfilled provided (\ref{aaaa}) is also satisfied.

Now, if we recall the origin of Dirac matrices in precanonical quantization as
the 
 quantum representations of differential forms, we can readily
 recognize the conditions
 (\ref{aaa}) and (\ref{aaaa}) as the inverse quantization map $q$ in Eq. (\ref{qmap})
 in the limit of
 the 
 infinitesimal elementary volume $\frac{1}{\ka}\rightarrow 0$,
 i.e. Eq. (\ref{aaaa})  is understood as follows:
 \beq \label{bekade}
\beta\ka\stackrel{q^{-1}}{\longmapsto}\delta(\mbox{\bf 0}) .
\eeq
Note, that one may think of the mapping in Eq. (\ref{bekade}) as the ``Wick rotation" in the hyperbolic complex plane $(1,\beta)$ combined with the limit $\ka\rightarrow\infty$.


\medskip

{\bf Fourth}, if Eq. (\ref{dtbpsi1}) is supposed to lead to a description
in terms of the wave functional $\BPsi$ alone,
then the third term in (\ref{delta-bpsi2}),
which is  proportional to $\der_y \Psi_\Sigma(\bx)$,
should cancel the  second term in  (\ref{dtbpsi1}),
which is also proportional to $\der_y \Psi_\Sigma(\bx)$.
This requirement leads to a condition which further
  restricts the dependence of
$\BPsi$ on  $\Psi_\Sigma(\bx)$ and $y(\bx)$, viz.,
\beq \label{dypsi}
\frac{\delta \BPsi}{\delta\Psi^T_\Sigma (\bx)}
 i\beta \gamma^i\der_iy(\bx)
 \mapsto -
\frac{\delta \deltab \BPsi}{\delta \Psi^T_\Sigma (\bx)
\deltab y(\bx)}~.  
\eeq
 By introducing the notation
 \beq \label{defphi}
\BPhi^{}({\bx})
:= \frac{\delta \BPsi}{\delta\Psi^T_\Sigma (\bx)}, ~
\eeq
and taking into account that $\der_i\delta(\mbox{\bf 0})=0$,
the solution of
Eq. (\ref{dypsi})
can be found in the form
\beq \label{bphi}
\BPhi^{}(\bx) =  \BXi([\Psi_\Sigma];\breve{\bx})~
e^{-iy(\bx)\gamma^i\der_iy(\bx)/\ka},
\eeq
where $\BXi([\Psi_\Sigma];\breve{\bx})$ denotes
 a functional of $\Psi_\Sigma (\bx')$
 at $\bx'\neq \bx$. Consequently,
\beq
\frac{\delta \BPhi (\bx )}{\delta \Psi^T_\Sigma (\bx )} =0
\eeq
or equivalently,
\beq \label{deltapsipsi}
\frac{\delta^2 \BPsi}{\delta\Psi_\Sigma (\bx)
 \otimes  \delta\Psi_\Sigma (\bx) } =0  .
\eeq
We note that
the latter equality is consistent with Eqs. (\ref{kade}) and  (\ref{aaa}).


Now,
Eqs. (\ref{defphi}) and (\ref{bphi}) lead to the following solution:
\beq \label{bpsi3}
\BPsi = 
 {\sf Tr}
\left \{\BXi([\Psi_\Sigma];\breve{\bx})~
e^{-iy(\bx)\gamma^i\der_iy(\bx)/\ka}~
 \frac{1}{\ka}\beta
 \Psi_\Sigma (\bx) \right \}_{\mbox{\Large $\rvert$} \scriptscriptstyle
 \beta\ka \stackrel{\mbox{\tiny $q^{-1}$}}{\mbox{$\longmapsto$}} \delta(\mbox{\bf\tiny 0})  },
\eeq
which is valid for any $\bx$
and
in combination with the inverse quantization map (\ref{bekade}).
It is easy to check that Eq. (\ref{bpsi3})
is  consistent with Eq. (\ref{v-term}).

 \medskip

The {\bf fifth} observation is 
that the last term in (\ref{delta-bpsi2}), evaluated on the solution (\ref{bpsi3}),  yields:
\beq
\frac{\deltab^2 \BPsi }{\deltab y(\bx)^2}
\mapsto
 (\nabla y(\bx))^2 \BPsi .
\eeq
Hence, it correctly reproduces the $\half (\nabla y(\bx))^2$ term in the
functional Schr\"o\-dinger equation (\ref{se}) with $\hat{\BH}$ given by  Eq. (\ref{hse}).
This calculation 
 thus indicates that those are the
  twisting
  phase factors
$e^{-iy(\bx)\gamma^i\der_iy(\bx)/\ka}$ in front of the precanonical wave functions
in (\ref{bpsi3}) which account for the non-ultralocality (in Klauder's terminology \cite{klauder})
of  relativistic scalar field theory.

\medskip

Our {\bf sixth} observation concerns  the first term in the right hand side of
Eq. (\ref{dtbpsi1}),  which contains the total derivative. Namely, by integration
by parts it takes the form
\beq \label{first}
\int \!\rd\bx\  {\sf Tr}\left \{
\left( i\frac{\rd}{\rd x^i}\BPhi (\bx) \right)
\gamma^i  \Psi_\Sigma  (\bx)
\right \} .
\eeq
Then, by taking the total derivative $\frac{\rd}{\rd x^i}$ of the explicit expression of
$\BPhi$ in Eq. (\ref{bphi}):
\beqa
\frac{\rd}{\rd x^i} \BPhi (\bx) &=& -\frac{i}{\ka} \BXi([\Psi_\Sigma];\breve{\bx})
e^{-iy(\bx)\gamma^i\der_iy(\bx)/\ka}
(\gamma^k\der_ky(\bx)\der_iy(\bx)
\nn \\
&+& y(\bx)\gamma^k \der_{ik}y(\bx) ) ,
\eeqa
and using the expression of $\BPsi$ in Eq. (\ref{bpsi3}), 
we transform Eq. (\ref{first}) to the following form: %
\beq \label{first0}
-i \BPsi\int \!\rd\bx\ (\gamma^k\der_ky(\bx)\der_iy(\bx)
+ y(\bx)\gamma^k \der_{ik}y(\bx) ) \gamma^i ,
\eeq
which obviously vanishes upon integration by parts.
Hence, the first (total derivative) term in the right hand side of (\ref{dtbpsi1})
does not contribute to the functional derivative
 equation describing the time evolution of  
 $\BPsi$.

\medskip

{\bf Finally,}
the functional $\BXi([\Psi_\Sigma(\bx)])$ in (\ref{bpsi3}) is specified
by combining all the above observations together and
noticing that the formula Eq. (\ref{bpsi3}) is valid for {\em any\/} $\bx$.
It can be accomplished only if the functional $\BPsi$ has the structure of the continuous
product of identical terms at all points $\bx$, viz.,
\beq\label{schrod}
\BPsi =
  {\sf Tr} \left \{\prod_\bx
e^{-iy(\bx)\gamma^i\der_iy(\bx)/\ka}
 \Psi_\Sigma (y(\bx), \bx, t)
\right \}_{\mbox{\Large $\rvert$}\scriptscriptstyle
 \beta\ka \stackrel{\mbox{$\scriptscriptstyle q^{-1}$}}{\mbox{$\longmapsto$}}
 \delta(\mbox{\tiny\bf{0}}) }.
\eeq

Thus we have obtained the expression of the Schr\"odinger wave functional in terms of
precanonical wave functions. The  equality in the
above expression implies the inverse of the Clifford algebraic
quantization map $q$ and the limit of the infinitesimal elementary volume element
$\frac{1}{\ka}\rightarrow 0$.
Moreover, the preceding consideration also derives term by term
the functional Schr\"odinger equation for the wave functional $\BPsi $
 from the precanonical Schr\"odinger equation for the wave function $\Psi$ restricted to the
subspace $\Sigma$.

As we have already mentioned in the previous paper \cite{myschrod11},
the inverse quantization map in the limit of
 infinitesimal $\frac{1}{\ka}$ means that
 \beq
 \frac{1}{\ka}\ \beta \stackrel{\mbox{$\scriptscriptstyle q^{-1}$}}{\mbox{$\longmapsto$}}
  \rd\bx.
 \eeq
  Therefore 
the expression of the wave functional in Eq. (\ref{schrod}) can be written in
the form of the multidimensional product integral
(c.f. \cite{productintegral}):  
\beq\label{schr-prod}
\BPsi = {\sf Tr}\left \{\prod_\bx
e^{-iy(\bx)\alpha^i\der_iy(\bx) \rd\bx}
  \Psi_\Sigma (y(\bx), \bx, t)_{\mbox{\large $\rvert$} \scriptscriptstyle    
  \frac{1}{\ka} \beta \mapsto \rd\bx }
\right \} ,
\eeq
which may be more practical to use than Eq. (\ref{schrod}).

\medskip

Further, let us recall that $\Psi_\Sigma $ obeys
Eq. (\ref{nsesigma}). According to  Eqs.  (\ref{first}) - (\ref{first0})
 the total derivative term does not contribute to the functional derivative
 equation
  for 
  $\BPsi$.
  In the case of scalar field theory,
  Eq. (\ref{nsesigma})
 without the total derivative term 
 (whose contribution to the Hamiltonian operator 
 vanishes, as it is shown in (\ref{first0})) 
 can be cast in the form
\beqa\label{mschr}
i\der_t\Psi_\Sigma &=& \frac{1}{2\ka} \beta \Big (i\ka \der_y + \gamma^i \der_i y(\bx)\Big)^2 \Psi_\Sigma  \nn \\
&&+\;\; \frac{1}{\ka} \beta \Big(V(y(\bx) + \frac12 (\nabla  y(\bx))^2\Big)\Psi_\Sigma
\;=:\; \beta \mathcal{E}\Psi_\Sigma .
\eeqa
The structure of
 the operator $\mathcal{E}$ in the
right hand side of Eq. (\ref{mschr}):
\beq \label{calE}
\mathcal{E} =
\frac{1}{2\ka}  \Big (i\ka \der_y + \gamma^i \der_i y(\bx)\Big)^2
+ \frac{1}{\ka}  \Big(V(y(\bx) + \frac12 (\nabla  y(\bx))^2\Big) ,
\eeq
resembles the structure of the magnetic Schr\"odinger operator
 in $y$-space
with the ``matrix magnetic potential" $\gamma^i \der_i y(\bx)$ and the
 ``electric potential"
$V(y(\bx)) + \frac12 (\nabla  y(\bx))^2$.

The ``magnetic" term in  Eq. (\ref{calE}) is pure gauge (in $y$-space),
 so that it does not change the eigenvalues of $\mathcal{E}$ in comparison with
 $\hat{H}$. Its influence reduces
to the phase shift of the eigenstates of $\hat{H}$ 
by the factor $e^{ i\gamma^i y(\bx)\der_i y(\bx) /\ka}$.
Note that Eq. (\ref{mschr}) is valid  in the fibers of field variables
and their first jets over each point $\bx$,
with  $\bx$-s  here just
 labeling the fiber
in which Eq. (\ref{mschr}) is written.

The addition $\frac12 (\nabla  y(\bx))^2$  to the ``electric potential" term
 in Eq. (\ref{calE})
 modifies the mass term in the potential term  of the DW Hamiltonian operator.
 Namely, by substituting it into Eq. (\ref{dtbpsi0})
 and integrating by parts using  the property (\ref{v-term}),
 we conclude that  under the restriction to $\Sigma$ the mass term
 $\frac12 m^2 y^2$  in $V(y)$  is replaced by
\beq
\frac12 y(\bx)(m^2 - \nabla^2) y(\bx)  .
\eeq
Correspondingly, the parameter $m$ in the expressions of precanonical wave functions
is formally replaced
 by $\sqrt{m^2 - \nabla^2}$,  when they are restricted to  $\Sigma$.

For example,
in the case of free massive scalar field theory the ground state of
the DW Hamiltonian operator,
$\hat{H} = - \half\ka^2\der_{yy} + \half m^2y^2$,
is given, up to the normalization factor, by
  $\Psi_0 \sim e^{-\frac{m}{2\ka}y^2}$,
   and its eigenvalue is $\half m\ka$ \cite{ka-qft,ka-qft2,ka-qft3}.
Then the eigenstates of $\beta \hat{H}$
 corresponding to the positive eigenvalues
 are
 given by $\sim\!(1+\beta) e^{-\frac{m}{2\ka}y^2}$.
Therefore,  the corresponding ground state wave function restricted to $\Sigma$: $\Psi_0{}_{\Sigma}$,
will take the form 
\beq \label{psi0sigma}
\Psi_{0\Sigma}
\sim e^{iy(\bx) \gamma^i\der_iy(\bx)/\ka} (1+\beta)
e^{-\frac{1}{2\ka}y(\bx) \sqrt{m^2 - \nabla^2}\,  y(\bx) } .
\eeq
\newcommand{\tinytexta}{\tiny
and it  fulfills  $\mathcal{E} \Psi_{0\, \Sigma}  = \omega \Psi_{0\, \Sigma} $
with the formal eigenvalue $\omega= \half\sqrt{m^2 - \nabla^2}$.
Note that the restriction to $\Sigma$ leads to the wave function
on the infinite jet space.
The solutions of Eq. (\ref{mschr}):
$$ i\der_t\Psi_\Sigma = \beta \mathcal{E}\Psi_\Sigma , $$
can be positive frequency
$\sim (1+\beta)\Psi_{0\Sigma}   e^{i \omega t }$ or
negative frequency $\sim (1-\beta)\Psi_{0\Sigma}   e^{-i \omega t }$.
...}
By substituting the last expression  into (\ref{schr-prod}) we
see that the ``magnetic" phase factors in Eq. (\ref{psi0sigma}) and Eq. (\ref{schrod})
will cancel each other,  so that  finally we
obtain
\beq \label{vac}
\BPsi \sim
  \mathsf{Tr}  \prod_\bx (1+\beta) e^{-\frac{1}{2}\beta y(\bx) \sqrt{m^2 - \nabla^2}  y(\bx)\rd\bx}
\sim e^{-\frac{1}{2}\int\!y(\bx) \sqrt{m^2 - \nabla^2}\,  y(\bx)\rd\bx} ,
\eeq
where the identity $\beta (1+\beta) = (1+\beta)$ is used (c.f. our earlier treatment in \cite{ik-pla}).

The right hand side of Eq. (\ref{vac}) reproduces the vacuum state solution
of the functional derivative Schr\"odinger equation for the free scalar field
(see e.g. \cite{hatfield}). Usually it
corresponds to the picture of the vacuum as the continuum of harmonic oscillators
with the zero-point energy $\half\sqrt{m^2+\bk^2}$ at every point of $\bk$-space.
Here the vacuum state of  free quantum scalar field is obtained as the
product of the ground state wave functions of the DW Hamiltonian operator
(which in this case corresponds to the  harmonic  oscillator in $y$-space)
 over all points $\bx$ of
  space.


\section{Conclusion}

Precanonical quantization, which is based on the space-time symmetric
generalization of the
Hamilto\-nian formalism to field theory (the De Donder-Weyl theory),
leads to the description of quantum fields in terms of
 Clifford-valued wave
functions on the bundle of field variables over space-time.
These wave functions obey a Dirac-like
generalization of the Schr\"odinger equation
 with the mass term replaced by the DW Hamiltonian operator.
The formulation
 introduces a small
parameter $\frac{1}{\varkappa}$ of the dimension of
 spatial volume, which appears on
dimensional grounds in the representation of precanonical quantum operators  
 and has the meaning of a minimal volume resolution.

A proper understanding of the connection between precanonical quantization and the standard
methods of quantization in field theory is important for the physical interpretation of
the results of precanonical quantization. In this paper, we  discuss how the results of
canonical quantization in the functional Schr\"odinger representation
are related to
 the
precanonical quantization and improve the arguments of
the previous discussions in \cite{ik-pla,myschrod11}.

Summarizing the considerations in Sect. 3 and those in
the preceding paper \cite{myschrod11}, we have proven that

\bigskip

{\sf Proposition: } {\em If\, $\Psi(y,x)$ is a precanonical wave function 
obeying the precanonical covariant
analogue of the Schr\"odinger equation {\rm (\ref{nse})}, 
 and \linebreak $\Psi_\Sigma(y(\bx),\bx,t)$
is its restriction to the subspace $\Sigma$ representing a field configuration
$y(\bx)$ at time $t$, then in the
limiting case $\beta\varkappa \mapsto \delta ({\mathbf{0}})$
or equivalently, $\frac{1}{\varkappa} \beta \mapsto \rd\bx$,
there exists a
 composite functional $\BPsi$ of\,  $\Psi_\Sigma$ and $y(\bx)$,
whose time evolution is governed
 by the standard functional derivative 
   Schr\"odinger equation 
   obtained from 
   canonical quantization. 
The time evolution of\, $\BPsi$ is completely determined by the time evolution of
$\Psi_\Sigma(y(\bx),\bx,t)$
 determined by the precanonical Schr\"odinger equation
restricted to $\Sigma$.
 The expression of the Schr\"odinger wave functional $\BPsi$ in terms of
  precanonical wave functions $\Psi$ is given by the product integral formula, Eq. {\rm (\ref{schr-prod})},
which is the {\sf necessary and sufficient} condition for the functional  $\BPsi$ to satisfy the
canonical Schr\"odinger equation 
 when $\Psi$ satisfies the precanonical Schr\"odinger equation.
 } 

\bigskip

This result leads to the conclusion that the canonical QFT
in the functional Schr\"odinger representation
is the limiting case of the theory obtained from precanonical quantization
corresponding to
$\beta\varkappa \mapsto \delta ({\mathbf{0}})$
 or   $\beta/ \varkappa \mapsto \rd\bx$, i.e. to the infinitesimal value of
 $\frac{1}{\ka}$.


 It is interesting that the
 introduction of the ultraviolet scale $\ka$ in precanonical quantization
does not modify the relativistic space-time at small distances. It rather defines the scale of
``very small" distances for the specific field theory under consideration.
 It is tempting, however, to interpret $\ka$ as the universal fundamental ultraviolet scale
 similar to the Planck scale, where the idea of the space-time continuum
 is supposed to break down due to quantum gravity effects.
 In this case, precanonical quantization might
 be able to provide new insights into Planck scale physics.


  Note in conclusion that the manifest respect for space-time symmetry within the precanonical quantization approach, together with the nonperturbative nature of the construction 
 in the case of interacting fields,  
potentially make it a suitable framework for the exploration of quantum gravity and quantum
gauge theories (c.f.  \cite{my-ym,my-gr} for the recent discussions). 

\bigskip

\noindent \small
{\bf Acknowledgments:} I am grateful to the University of Wuppertal, the University of G\"ottin\-gen and the Erwin Schr\"odinger Institute in Vienna for their financial support in helping me participate in the Workshops they have hosted in 2013.  I am thankful to J. Akram whose remarks on an earlier version of the manuscript have helped me to improve the presentation. Thanks are due to  H. Kleinert, A. Pelster and J. Dietel for their kind hospitality at the Research Center of Einstein Physics and to J. Kouneiher and M. Wright for their comments and suggestions on a later version of the manuscript.

\address{National Quantum Information Centre in Gda\'nsk (KCIK)\\
81-824 Sopot, Poland
\medskip \\
Research Center of Einstein Physics, Free University of Berlin,\\
14195 Berlin, Germany
\medskip \\
\email{kanattsi@gmail.com}}

\end{document}